%
% the following is to use blackboard bold fonts --
\let\useblackboard=\iftrue
%
% activate this if you don't have them.
%\let\useblackboard=\iffalse
%
% You might also need to remove this line.
\newfam\black
\input harvmac.tex
\def\Title#1#2{\rightline{#1}
\ifx\answ\bigans\nopagenumbers\pageno0\vskip1in%
\baselineskip 15pt plus 1pt minus 1pt
\else%\special{papersize=11in,8.5in}%
\def\listrefs{\footatend\vskip 1in\immediate\closeout\rfile\writestoppt
\baselineskip=14pt\centerline{{\bf References}}\bigskip{\frenchspacing%
\parindent=20pt\escapechar=` \input
refs.tmp\vfill\eject}\nonfrenchspacing}
\pageno1\vskip.8in\fi \centerline{\titlefont #2}\vskip .5in}

\ifx\answ\bigans\def\tcbreak#1{}\else\def\tcbreak#1{\cr&{#1}}\fi
\useblackboard
\message{If you do not have msbm (blackboard bold) fonts,}
\message{change the option at the top of the tex file.}
\font\blackboard=msbm10 scaled \magstep1
\font\blackboards=msbm7
\font\blackboardss=msbm5
%\newfam\black
\textfont\black=\blackboard
\scriptfont\black=\blackboards
\scriptscriptfont\black=\blackboardss

\else

\fi
\def\yboxit#1#2{\vbox{\hrule height #1 \hbox{\vrule width #1
\vbox{#2}\vrule width #1 }\hrule height #1 }}
\def\fillbox#1{\hbox to #1{\vbox to #1{\vfil}\hfil}}
\def\ybox{{\lower 1.3pt \yboxit{0.4pt}{\fillbox{8pt}}\hskip-0.2pt}}
\def\np#1#2#3{Nucl. Phys. {\bf B#1} (#2) #3}
\def\pl#1#2#3{Phys. Lett. {\bf #1B} (#2) #3}

\def\physrev#1#2#3{Phys. Rev. {\bf D#1} (#2) #3}

\def\comments#1{}

\def\Tr{{{\rm Tr\  }}}

\def\CM{{\cal M}}

\def\II{\relax{I\kern-.07em I}}

\def\IZ{\relax\ifmmode\mathchoice
{\hbox{\cmss Z\kern-.4em Z}}{\hbox{\cmss Z\kern-.4em Z}}
{\lower.9pt\hbox{\cmsss Z\kern-.4em Z}}
{\lower1.2pt\hbox{\cmsss Z\kern-.4em Z}}\else{\cmss Z\kern-.4em
Z}\fi}
\def\IB{\relax{\rm I\kern-.18em B}}
\def\IC{{\relax\hbox{$\inbar\kern-.3em{\rm C}$}}}
\def\ID{\relax{\rm I\kern-.18em D}}
\def\IE{\relax{\rm I\kern-.18em E}}
\def\IF{\relax{\rm I\kern-.18em F}}
\def\IG{\relax\hbox{$\inbar\kern-.3em{\rm G}$}}
\def\IGa{\relax\hbox{${\rm I}\kern-.18em\Gamma$}}
\def\IH{\relax{\rm I\kern-.18em H}}
\def\II{\relax{\rm I\kern-.18em I}}
\def\IK{\relax{\rm I\kern-.18em K}}
\def\IP{\relax{\rm I\kern-.18em P}}
%\def\IX{\relax{\rm X\kern-.01em X}}
%this doesn't work

\font\cmss=cmss10 \font\cmsss=cmss10 at 7pt
\def\IR{\relax{\rm I\kern-.18em R}}

\def\BR{\IR}

\def\IZ{\relax\ifmmode\mathchoice
{\hbox{\cmss Z\kern-.4em Z}}{\hbox{\cmss Z\kern-.4em Z}}
{\lower.9pt\hbox{\cmsss Z\kern-.4em Z}}
{\lower1.2pt\hbox{\cmsss Z\kern-.4em Z}}\else{\cmss Z\kern-.4em
Z}\fi}
\def\IB{\relax{\rm I\kern-.18em B}}
\def\IC{{\relax\hbox{$\inbar\kern-.3em{\rm C}$}}}
\def\ID{\relax{\rm I\kern-.18em D}}
\def\IE{\relax{\rm I\kern-.18em E}}
\def\IF{\relax{\rm I\kern-.18em F}}
\def\IG{\relax\hbox{$\inbar\kern-.3em{\rm G}$}}
\def\IGa{\relax\hbox{${\rm I}\kern-.18em\Gamma$}}
\def\IH{\relax{\rm I\kern-.18em H}}
\def\II{\relax{\rm I\kern-.18em I}}
\def\IK{\relax{\rm I\kern-.18em K}}
\def\IP{\relax{\rm I\kern-.18em P}}
%\def\IX{\relax{\rm X\kern-.01em X}}
%this doesn't work

\font\cmss=cmss10 \font\cmsss=cmss10 at 7pt
\def\IR{\relax{\rm I\kern-.18em R}}

\def\bS{{\bf S}}

\def\tilde{\widetilde}
\def\frac#1#2{{{#1} \over {#2}}}

\Title{ \vbox{\baselineskip12pt\hbox{hep-th/9705221}
\hbox{RU-97-42}}}
{\vbox{\centerline{New Theories in Six Dimensions and}
\centerline{}
\centerline{Matrix Description of M-theory on $T^5$ and $T^5/Z_2$}}} 

\centerline{Nathan Seiberg}
\smallskip
\smallskip
\centerline{Department of Physics and Astronomy}
\centerline{Rutgers University }
\centerline{Piscataway, NJ 08855-0849}
\bigskip
\bigskip
\noindent
We present four infinite series of new quantum theories with
super-Poincare symmetry in six dimensions, which are not local quantum
field theories.  They have string like excitations but the string
coupling is of order one.  Compactifying these theories on $T^5$ we find
a Matrix theory description of M theory on $T^5$ and on $T^5/\IZ_2$,
which is well defined and is manifestly U-duality invariant.

%\draftmode
\Date{May 1997}

\newsec{Introduction}

The Matrix theory 
\ref\bfss{T. Banks, W. Fischler, S. Shenker and L. Susskind,
hep-th/9610043, \physrev{55}{1997}{112}.}
description of
compactification of M theory on $T^3$ involves a supersymmetric gauge
theory in 3+1 dimensions
\nref\tori{W. Taylor IV, hep-th/9611042, \pl{394}{1997}{283}.}%
\nref\sus{L. Susskind, hep-th/9611164.}%
\nref\fhrs{W. Fischler, E. Haylo, A. Rajaraman and L. Susskind,
hep-th/9703102.}%
\refs{\bfss - \fhrs}.  Compactification on higher dimensional tori
seems to involve higher dimensional gauge theories which are not
renormalizable.  Therefore, this prescription must be modified.
Indeed, in
\nref\rozali{M. Rozali, hep-th/9702136.}%
\nref\brs{M. Berkooz, M. Rozali and N. Seiberg, hep-th/9704089.}%
\refs{\rozali,\brs} it was suggested to define the compactification on
$T^4$ in terms of a six dimensional field theory with (2,0)
supersymmetry.  This field theory is at a fixed point of the
renormalization group
\nref\wittentwoz{E. Witten, hep-th/9507121, Contributed to STRINGS 95:
Future Perspectives in String Theory, Los Angeles, CA, 13-18 Mar 
1995.}%
\nref\andy{A. Strominger, hep-th/9512059, \pl{383}{1996}{44}.}%
\nref\wittenbranes{E. Witten, hep-th/9512219, \np{463}{1996}{383}.}% 
\nref\sixteen{N. Seiberg, hep-th/9705117.}%
\refs{\wittentwoz - \sixteen}.  Trying to extend this discussion to
compactification on $T^5$ has led the authors of \brs\ to conjecture the
existence of a new theory (parametrized by an integer $N$) with six
dimensional super-Poincare symmetry, which has string like excitations
and T-duality.  Having string like excitations does not necessarily mean
that the theory is not a local quantum field theory.  For example, the
(2,0) theory has string like excitations, which become tensionless at
the critical point, but the theory there appears to be a local quantum
field theory.  However, the existence of T-duality in our theory shows
that the underlying geometry of the base on which the theory ``lives''
is ambiguous.  Therefore, the theory does not have a standard energy
momentum tensor, and it is not a local quantum field theory.

One of the purposes of this paper is to give an existence proof for this
theory, and to explore some of its properties.  Our analysis will lead
us to find several such theories.

Our strategy will be to assume that M theory exists, and to study
various 5-branes in the theory.  We will focus on various string
theory limits of M theory, and will explore the limits as the string
coupling, $g_s$, goes to zero.  In these limits, the modes in the bulk
of space-time decouple.  However, the modes on the branes remain
interacting.  The 5-branes are singular objects, and therefore even in
these limits the theory on the 5-branes is still non-trivial.

We find four infinite classes (labeled by the number of
parallel 5-branes, $N$) of new theories.  Two of them are obtained from
the NS5-branes of the type IIA and the type IIB string theories.  The
other two are found on 5-branes (instantons) of the $Spin(32)$ and
the $E_8 \times E_8$ heterotic string theories.

Upon compactification on tori these theories inherit the T-duality
of the underlying string theory.  Two facts are crucial in establishing
that T-duality acts on these theories.  First, $g_s=0$ is a fixed point
of T-duality.  Second, wrapped NS5-branes remain wrapped 
NS5-branes after T-duality.  In this respect they differ from D-branes
\ref\joe{J. Polchinski, S. Chaudhuri, and C. Johnson, hep-th/9602052; J.
Polchinski, hep-th/9611050.}.
If these had been ordinary local quantum field theories, the base on
which the fields propagate would have been unambiguous.  Its geometry
would have been related to the existence of a well defined energy
momentum tensor.  Here, the existence of T-duality makes the base
ambiguous, and therefore these theories do not have a unique energy
momentum tensor.  Instead, there are several different operators which
can be interpreted as the energy momentum tensor.  In different limits
of the parameters different operators are naturally identified as the
energy momentum tensor.  (Note that this is unlike the situation in
theories of gravity or in topological field theories, where there is
no energy momentum tensor at all.)  We conclude that our new theories
are not local quantum field theories.

We should comment that somewhat related ideas were discussed in the
context of black holes in
\ref\dvv{R. Dijkgraaf, E. Verlinde and H. Verlinde, hep-th/9603126,
\np{486}{1997}{77}; hep-th/9604055, \np{486}{1997}{89}.}.
The precise relation between this work and our proposal is not
completely clear to us.

In section 2 we present two new classes of theories with 16 supercharges
in six dimensions.  One of them has (1,1) supersymmetry and its low
energy behavior is a $U(N)$ gauge theory with 16 supercharges.  The
other has (2,0) supersymmetry, and its low energy behavior is an
interacting field theory.  They are continuously connected upon
compactification on a circle.  Compactifying this theory on $T^5$ we
find a well defined and manifestly U-duality invariant description of M
theory on $T^5$.

In section 3 we present two new classes of theories with (1,0)
supersymmetry in six dimensions.  One of them has a $Spin(32)$ global
symmetry.  At low energies it becomes an $SP(N)$ gauge theory with 16
hypermultiplets in the fundamental representation and one hypermultiplet
in the antisymmetric tensor representation.  The other has $E_8 \times
E_8$ global symmetry and is an interacting field theory at low
energies.  They are continuously connected upon compactification on a
circle.  Compactifying this theory on $T^5$ we find a well defined and
manifestly U-duality invariant description of M theory on $T^5/\IZ_2$.

In section 4 we mention some extensions of our work and suggest the
existence of other new theories.

\newsec{New Theories with 16 supercharges}

\subsec{The Theories}

Here we consider the theory of solitonic 5-branes in the two type II
theories.

We start with $N$ NS5-branes in the type IIB theory.  Using S-duality we
can relate this to $N$ D5-branes.  These have a $U(N)$ gauge symmetry
\ref\bound{E. Witten, \np{460}{1996}{335}, hep-th/9510135.}
and gauge coupling 
\eqn\dgauge{{1 \over g_{D}^2 }= {M_s^2 \over g_s},}
where $M_s$ is the string scale.  In the weak coupling limit the gauge
theory becomes weakly coupled.  This means that the gauge theory on the
D5-branes breaks down at energies of order $M_s/\sqrt{g_s}$ and needs new
dynamics at that scale.
Using S-duality, the gauge coupling in the $U(N)$ gauge theory on the
NS5-branes is 
\eqn\nsgauge{{1 \over g_{NS}^2 }= M_s^2.}
We see that even when $g_s=0$ the gauge coupling does not vanish.  The
gauge theory on the NS5-branes breaks down at energies of order $M_s$ and
needs new degrees of freedom there.  Since the underlying string theory
makes sense, we expect that the theory of the NS5-branes also makes
sense.  It includes the necessary degrees of freedom at energies of
order $M_s$ to yield a consistent theory.  It is crucial to stress that
the theory we are left with is not the full underlying string theory.
The latter has vanishing coupling constant and most of its modes
decouple. 

We do not have a complete description of the theory at energies of order
$M_s$.  However, using the low energy gauge theory we can detect that it
includes strings whose tension is $M_s^2$.  The gauge theory has
instantons which are strings in six dimensions.  Their tension is ${1
\over g_{NS}^2} = M_s^2$.  They can be interpreted as fundamental
strings within the NS5-branes.  Since $g_s=0$, they cannot leave the
branes.  Similar instantons in D5-branes were studied in
\ref\braneswith{M.R. Douglas, hep-th/9512077.}.
The detailed properties of these strings depend on the details of the
theory which we do not know. 

We can repeat this analysis for $N$ NS5-branes in the IIA theory.  Here
we find at low energies a six dimensional theory with (2,0)
supersymmetry.  The moduli space of vacua is
\eqn\modtwozero{\CM={(\BR^4 \times S^1)^N \over \bS_N}.}
The circle $S^1$ originates from the circle which relates M theory to
the IIA theory.  As in \sixteen, we take the moduli, $\Phi$, to be
fields of dimension 2.  In this normalization their kinetic terms have
no dimensionful coefficient and the circumference of the $S^1$ factor is
$M_s^2$.  At the singularities of the moduli space \modtwozero\ we find
the interacting (2,0) theory associated with the group $U(N)$ (for a
review, see \sixteen).  At the vicinity of these singularities the
theory includes strings, whose tension vanishes at the singularities.
Clearly, the full theory is not just this field theory.  At energy of
order $M_s$ new degrees of freedom become important. Some of them can be
identified from an M theory point of view.  M theory membranes which
wind around the circle and end on the 5-branes appear as strings whose
tension is $M_s^2$.  As before, they can be identified as the
fundamental strings which cannot leave the 5-branes because $g_s=0$.

Consider now the compactification of the underlying type II string
theory on a circle and wrap the NS5-branes on this circle.  T-duality on
this circle is not a symmetry.  It relates a compactification of the IIA
theory on a circle with radius $\Sigma^A$ with a compactification of the
IIB theory on a circle of radius $\Sigma^B={1 \over M_s^2 \Sigma^A}$.
The theory in the non-compact directions is a five dimensional $U(N)$
gauge theory with coupling constant
\eqn\fivedg{{1 \over g_5^2 } = \Sigma^B M_s^2 = {1 \over \Sigma^A}.}
The moduli space of vacua is as in \modtwozero.  From the IIA point of
view we define the scalars $\phi = \Sigma^A \Phi$.  From the IIB point
of view four of the scalars are as in the higher dimensional theory and
the fifth originates from the gauge fields.  Its periodicity is ${1
\over \Sigma^B} = \Sigma^A M_s^2$.

It is also illuminating to compare some of the different BPS states in
the five dimensional theory.  For example, the ``W-bosons,'' which are
obvious in the low energy IIB theory, arise as strings winding around
$\Sigma^A$ in the IIA theory.  The momentum modes of the IIA theory with
masses $ n \over \Sigma^A$ correspond to strings with tension $M_s^2$
(instantons) winding $n$ times around $\Sigma^B$.  Similarly, momentum
modes of the IIB theory with masses $n \over \Sigma^B$ correspond to
strings with tension $M_s^2$ winding $n$ times around $\Sigma^A$.  This
is the standard exchange of momentum and winding modes in T-duality done
here for the modes on the 5-branes.  It is an amusing exercise to compare
also the strings in the five dimensional gauge theory.

We can continue further to compactify these theories on $T^5$.  The data
of the compactification are the metric and the $B$ field on the $T^5$
(the RR fields of the string theory decouple at $g_s=0$).  The
$B$ field couples to the strings in these theories.  In terms of 
the gauge theory of the six dimensional IIB theory these strings are
instantons and therefore the coupling is\foot{We thank E. Witten for a
useful discussion on this point.} $\int B \wedge \Tr F \wedge F$.  The
total number of parameters in the compactification is 25 and they
parametrize 
\eqn\sofifi{SO(5,5,\IZ) \backslash SO(5,5)/(SO(5) \times SO(5)),}
which is the standard Narain space inherited from the underlying string
theory.  The full duality group $SO(5,5,\IZ) $ follows as in string
theory. 

\subsec{Some Excitations}

Some of the excitations of these theories will be useful below.
Consider a bound state of $N$ NS5-branes and $n$ D$p$-branes ($p$ even
for IIA and $p$ odd for IIB). For simplicity we consider a
compactification on a torus with right angles and no $B$ field.  The
energy of such a state in the $g_s=0$ limit is
\eqn\eners{\eqalign{ E& = \lim_{g_s \rightarrow 0} \sqrt {
\left({N\Sigma_1 \Sigma_2 \Sigma_3 \Sigma_4 \Sigma_5 M_s^6 \over
g_s^2}\right)^2 + \left({n\Sigma_{i_1}...\Sigma_{i_p}M_s^{p+1 } \over
g_s}\right)^2} - {N\Sigma_1 \Sigma_2 \Sigma_3 \Sigma_4 \Sigma_5 M_s^6
\over g_s^2} \cr 
&={(n\Sigma_{i_1}...\Sigma_{i_p})^2 \over 2
N M_s^{4-2p} \Sigma_1 \Sigma_2 \Sigma_3 \Sigma_4 \Sigma_5}, }}
where the factors of $g_s$ are determined by the tensions of the objects
-- $1 \over g_s^2$ for the NS5-branes and $1 \over g_s$ for the
D5-branes and the $\Sigma_i$'s are $\Sigma^A_i$ or $\Sigma^B_i$
depending on the description we use (NS5-branes in IIA or in IIB).
Note that these states have finite energy for $g_s=0$.
Clearly, the T-duality group combines the wrapped even D-branes of the
IIA theory (the even homologies) to a $\bf 16$ of
$SO(5,5)$ and the wrapped odd D-branes to $\bf 16'$ of $SO(5,5)$.

The states with energies \eners\ can be identified in terms of
excitations of our two low energy theories.  They are characterized by
carrying fluxes of the $U(1)$ part of the low energy gauge theories.
Near the limit where the description as NS5-branes in IIA is
appropriate all the fluxes are visible as in \brs.  To keep the
notation uniform we dualize the compact scalar to a four form gauge
field, $A^{(4)}$, with a five form field strength, $F^{(5)}$, whose
kinetic term is multiplied by $1 \over M_s^4$.  This field has one
magnetic flux and five electric fluxes.  The energies of states
carrying $n$ units of these fluxes are easily computed using the free
Lagrangian
\eqn\energfa{\eqalign{
&E^M={n^2 \over 2NM_s^4 \Sigma_1^A \Sigma_2^A \Sigma_3^A \Sigma_4^A
\Sigma_5^A} \cr
&E^E_i= { M_s^4 \Sigma_1^A \Sigma_2^A \Sigma_3^A \Sigma_4^A
\Sigma_5^A n^2 \over 2N(\Sigma_i^A)^2 }, \cr}}
where we have set the normalization (factor of 2) to agree with \eners.
In addition, there are ten fluxes of the self-dual $H$ field.  The
energies of the corresponding states are
\eqn\energfh{E^H_{ij} = {(n\Sigma_i^A \Sigma_j^A)^2 \over 2N\Sigma_1^A
\Sigma_2^A \Sigma_3^A \Sigma_4^A \Sigma_5^A} }
where again, we have set the normalization to agree with \eners.
These can be identified with the 16 states in \eners\ corresponding to
bound states of $N$ NS5-branes with $n$ D0-branes ($E^M$), $n$ D2-branes
($E_{ij}^H$), or $n$ D4-branes ($E_i^E$).

Near the region of the parameters where the light degrees of freedom are
those of the (1,1) super-Yang-Mills theory the states can be identified
as in \fhrs.   Here we find 15 fluxes: five electric and ten
magnetic fluxes.  States with $n$ units of flux have energies
\eqn\energf{\eqalign{
&E^E_i= {(n\Sigma_i^B)^2\over  2NM_s^2 \Sigma_1^B \Sigma_2^B \Sigma_3^B
\Sigma_4^B \Sigma_5^B }\cr
&E^M_{ij} = {M_s^2 \Sigma_1^B \Sigma_2^B \Sigma_3^B \Sigma_4^B \Sigma_5^B
n^2 \over 2N (\Sigma_i^B \Sigma_j^B)^2}, \cr}}
where again we have set the normalization to agree with \eners.  The
bound states of $N$ NS5-branes and $n$ D1-branes correspond to the
states with electric flux ($E^E_i$) while the bound states with $n$
D3-branes correspond to the states with magnetic flux ($E_{ij}^M$).  We
can also use the T-duality transformation between the two
descriptions, $\Sigma_{1,2,3,4}^B=\Sigma_{1,2,3,4}^A$, $\Sigma_5^B= {1
\over M_s^2 \Sigma_5^A}$ to match with the states of \energfa\ and
\energfh.  The remaining  16'th ``missing'' state has energy
\eqn\energnf{E = {n^2 \over 2N} M_s^6 \Sigma_1^B \Sigma_2^B \Sigma_3^B
\Sigma_4^B \Sigma_5^B.}
It corresponds to a bound state of $N$ NS5-branes and $n$ D5-branes
\eners.  It has finite energy density, $n^2M_s^6/2N$.  Therefore, it
corresponds to another sector of the Hilbert space of the infinite
volume theory.  This explains why it cannot be identified as an
excitation in the low energy theory around the ordinary vacuum -- in the
super-Yang-Mills theory.

\subsec{Matrix Theories of M Theory}

We can now use these theories as Matrix theories of M theory as
suggested in \brs.  Consider M theory compactified on $T^5$ with sizes
$L_i$ ($i=1,...,5$).  For simplicity we take all the angles to be right
angles and we set the three form field to zero.  To describe this theory
we take $N$ NS5-branes in IIB with scale $M_s$ on $T^5$ with sizes
$\Sigma_i^B$ such that
\eqn\montf{\eqalign{
&\Sigma_i^B = {l_p^3 \over R L_i} \cr
&M_s^2 = {R^2 L_1L_2L_3L_4L_5 \over l_p^9}, \cr}}
where $R$ is the length of the longitudinal direction and $l_p$ is the
eleventh dimensional Planck length.  Alternatively, we can use $N$
NS5-branes of the IIA theory with
\eqn\montfa{\eqalign{
&\Sigma_{1,2,3,4}^A = {l_p^3 \over R L_{1,2,3,4}} \cr
&\Sigma_5^A = {l_p^6 \over R L_1L_2L_3L_4} \cr
&M_s^2 = {R^2 L_1L_2L_3L_4L_5 \over l_p^9} .\cr}}
In some range of parameters the first description becomes approximately
a six dimensional gauge theory which coincides with the prescription of
\fhrs.  It is crucial to stress that this gauge theory is only an
approximate description in some range of parameters;  the
complete theory is that of the NS5-branes.  The second description
with the parameters in \montfa\ matches that of \brs.

As explained in \brs, already for the case of compactification on
$T^4$, there is no unique way to extract the space-time geometry from
these theories.  The point is that space-time is the moduli space of
vacua of the theory.  However, in quantum mechanics there is no moduli
space of vacua -- we have to integrate over all vacua.  The closest to
moduli space of vacua is a situation, as in the Born-Oppenheimer
approximation, where some of the energy levels are much lower than
others and almost form a continuum.  Then we can describe them as
degrees of freedom moving on some space -- ``the moduli space of
vacua'' --
which is identified as the space-time of M theory.  Already for the
case of M theory on $T^4$, this procedure was shown to be ambiguous;
for different values of the space-time moduli (parameters in the
quantum mechanics) we find different natural space-time
interpretations \brs.  Here, because of the T-duality, we see a
phenomenon which is more subtle.  Even the underlying base space on
which the new theory propagates is ambiguous.

In addition to the 16 supercharges $Q$ of our new theories there are
also 16 non-linearly realized supercharges $\tilde Q$, which together
give the 32 supercharges of M theory.  This split between them is
standard in the light-cone frame.  As in
\ref\bss{T. Banks, N. Seiberg and S.H. Shenker, hep-th/9612157,
\np{490}{1997}{91}.},
we write the supersymmetry algebra in the light-cone frame including the
central charges.  We set to zero all the central charges which are not
scalars of the transverse space rotation group $SU(2) \times SU(2)$.
This symmetry appears as an R-symmetry of our theories (the low energy
theory on the NS5-brane in IIA has $SP(2)$ R-symmetry, which is broken at
the scale $M_s$ to an $SU(2) \times SU(2)$ subgroup).  Together with the
space rotation group the symmetry is $SU(2) \times SU(2) \times
Spin(5)$.  The supercharges transform under it as two copies of $({\bf
2,1,4}) \oplus ({\bf 1,2,4})$.  The supersymmetry algebra is
\eqn\susylc{\eqalign{
&\{Q_\alpha^i, Q_\beta^j\}=2H \epsilon_{\alpha \beta} J^{ij} +
\epsilon_{\alpha \beta} \Gamma_I^{ij} Z^I \cr
&\{Q_{\dot \alpha}^i, Q_{\dot \beta}^j\}=2H \epsilon_{\dot \alpha \dot
\beta} J^{ij} + \epsilon_{\dot \alpha \dot \beta} \Gamma_I^{ij} \tilde
Z^I \cr 
&\{Q_\alpha^i, \tilde Q_\beta^j\}=\epsilon_{\alpha \beta} Z^{ij} \cr
&\{Q_{\dot \alpha}^i, \tilde Q_{\dot \beta}^j\}=\epsilon_{\dot \alpha
\dot \beta} \tilde Z^{ij} \cr 
&\{\tilde Q_\alpha^i, \tilde Q_\beta^j\}= 2 P^+ \epsilon_{\alpha \beta}
J^{ij} \cr
&\{\tilde Q_{\dot\alpha}^i, \tilde Q_{\dot \beta}^j\}= 2 P^+
\epsilon_{\dot \alpha \dot \beta} J^{ij} \cr}} 
(all other anticommutators vanish),
where $\alpha,\beta =1,2$ and $\dot \alpha, \dot \beta = 1,2$ label the
spinors of the two $SU(2)$ factors, and $i,j=1,...,4$, $I=1,...,5$ and
$J^{ij}$  are the spinor indices, the vector index and the invariant
tensor of the $Spin(5)$ space rotation group. 

The central charges $Z^I,\tilde Z^I, Z^{ij}, \tilde Z^{ij}$ in \susylc\
can be interpreted both in our Matrix theory and in the space-time M
theory.  Inspection of \susylc\ shows that if $Z^I$ or $\tilde Z^I$ are
not zero, and we rescale them by a constant $n$, then the energy of the
BPS states is proportional to $n$.  Therefore, if many states which are
labeled by $n$ become light, we recognize the dispersion relation of
relativistic particles (energy proportional to the momentum).  We
interpret these central charges as momenta in our theory.  From the type
II string point of view $Z^I$ and $\tilde Z^I$ are the left and right
moving momenta of the string.  In various limits of the Narain moduli
space, some of the excitations carrying these quantum numbers become
light and we can interpret them as ordinary momenta.  Again, the fact
that we have 10 such momenta rather than 5 shows that these are not
theories on a well defined five torus.

The space-time interpretation of $Z^I$ and $\tilde Z^I$ is the winding
numbers of strings stretched along the longitudinal directions.  This
interpretation can be derived using the full underlying Lorentz symmetry
(rather than just the transverse part) as in \bss.  Alternatively, this
follows from the fact that their energy, which is interpreted as $P^-$ in
space-time, is proportional to their charge \bss.  As a consequence of
this interpretation it is clear that $Z^I$ and $\tilde Z^I$ should be
proportional to the length of the longitudinal direction $R$.  Indeed,
we see in \montf\ and \montfa\ that the lengths scale like $1 \over R$
and therefore the momenta scale like $R$.

It is nice to check this interpretation of longitudinal strings with
the momenta of the underlying Matrix theory in simpler contexts.
For compactifications on $S^1$ the momentum in the 1+1 dimensional
gauge theory was interpreted in
\nref\bs{T. Banks and N. Seiberg, hep-th/9702187.}%
\nref\strt{R. Dijkgraaf, E. Verlinde and H. Verlinde,
hep-th/9703030.}% 
\refs{\bs,\strt} as the winding number of longitudinal strings.  In the
context of compactifications on $T^2$ and the IIB string theory in ten
dimensions  
\nref\ss{S. Sethi and L. Susskind, hep-th/9702101.}%
\refs{\ss,\bs} the momenta along the two space dimensions were
identified as the winding numbers of longitudinal NS and D-strings \bs. 
A new element appears in compactifications on $T^4$.  Here, a 4-brane
wrapping the $T^4$ can lead to a longitudinal string.  Its winding
number was identified in
\nref\grw{O.J. Ganor, S. Ramgoolam and W. Taylor IV, hep-th/9611202.}%
\refs{\grw,\bss} as the instanton number of the $4+1$ dimensional gauge
theory.  Precisely this object was identified in \rozali\ as the
momentum along the fifth direction in the matrix description of this
theory.

We now turn to the interpretation of the central charges $Z^{ij}$ and
$\tilde Z^{ij}$ in \susylc.  From the space-time picture it is
clear that for particles (0-branes) $Z^{ij}=\tilde Z^{ij}$ while for
4-branes in the transverse directions $Z^{ij}=-\tilde Z^{ij}$.  The
latter are rather singular objects and therefore we set $Z^{ij}=\tilde
Z^{ij}$.  Rescaling the values of the charges of a state by $n$, we
see from \susylc\ that the energy $H=P^-$ scales like $n^2 \over P^+$
which is consistent with the interpretation of these objects as
particles.  They correspond to one state of a wrapped 5-brane, ten
states of wrapped 2-branes  and five momentum modes.  From the Matrix
point of view these charges are the fluxes 
discussed above.  States with nonzero charges are the bound states of
our system with D-branes.

\newsec{New Theories with 8 supercharges}

In this section we construct two more theories starting with the
5-branes of the two heterotic theories at zero coupling.  These two
theories have (1,0) super-Poincare symmetry in six dimensions.  Most of
the conceptual issues are similar to the discussion in the type II
theory and will not be repeated here.

One theory, based on $N$ $Spin(32)$ instantons has as its low energy
theory an $SP(N)$ gauge theory with hypermultiplets in the antisymmetric
tensor of $SP(N)$ and 16 fundamentals
\ref\witins{E. Witten, hep-th/9511030, \np{460}{1995}{541}}.
The global symmetry of this theory is $Spin(32)$.  Its gauge coupling is
${1 \over g^2 }= M_s^2$.  Again, it is important that $g$ does not
vanish for $g_s=0$.  The theory includes string like excitations, which
can be interpreted as $SP(N)$ instantons.  Their tension is ${1 \over
g^2 }= M_s^2$ and they can be interpreted as the fundamental heterotic
strings \braneswith. Their detailed properties depend on the structure of
the theory at energies of order $M_s$.  Note that unlike \witins, we
propose a complete theory and not just an effective description at low
energies.

The other new theory is based on instantons in the $E_8 \times E_8$
heterotic string.  Here the low energy theory is an interacting quantum
field theory, which has string like excitations
\ref\ees{O. Ganor and A. Hanany,  hep-th/9602120, \np{474}{1996}{122};
N. Seiberg and E. Witten, hep-th/9603003, \np{471}{1996}{121}; M. Duff,
H. Lu and C.N. Pope, hep-th/9603037, \pl{378}{1996}{101}.}.
For $N$ such instantons the moduli space of the full theory is
\eqn\modee{{(\BR^4 \times (S^1/\IZ_2))^N \over {\bf S}_N}.}
The scalars in this theory have dimension 2 such that their kinetic
terms are dimensionless.  The size of the $S^1/\IZ_2$ factor is
$M_s^2$. 

As in the type II theory, upon compactification on a circle these two
six dimensional theories become the same.  Compactification on $T^5$
depends on 105 parameters in
\eqn\hetmod{SO(21,5,\IZ) \backslash SO(21,5) /(SO(21)\times SO(5)).}
As in the type II theory, the T-duality is inherited from the heterotic
string and implies that these are not local quantum field theories.

We now use these theories as Matrix theories for M theory.  We should
find a six dimensional theory with (2,0) space-time supersymmetry.  This
leads us to guess that the answer is the compactification of M theory on 
$T^5/\IZ_2$, which is the same as the IIB theory on K3
\ref\tbkt{K. Dasgupta and S. Mukhi, hep-th/9512196, \np{465}{1996}{399};
E. Witten, hep-th/9512219, \np{463}{1996}{383}.}.
As a first indication that this is the right answer we recognize the
moduli space \hetmod\ as the moduli space of vacua of this space-time
theory.  The local structure of the moduli space is fully determined by
supersymmetry and the lack of space-time anomalies (these determine the
number 21).  The fact that the global structure is correct is less
trivial. 

The identification can be made more precise by going to the limit, where
we can use the $SP(N)$ gauge theory compactified on $T^5$.  There we can
find the low energy modes, and identify the moduli space of the quantum
mechanical system, which can be interpreted as space-time.  As in
\ref\spkthr{A. Fayyazudin and D.J. Smith, hep-th/9703208; N. Kim and
S.-J. Rey, hep-th/9705132.}, 
one branch of the moduli space is $(T^5/\IZ_2)^N/{\bf S}_N$, which
corresponds to $N$ zero branes moving on $T^5/\IZ_2$.  This fact
completes the identification of this compactification.

We can repeat the analysis in the type II theory and find the 26 momenta
in our new theories as central charges in the space-time supersymmetry
algebra.  They correspond to longitudinal strings.  Unlike the type II
theory, here there are no D-branes and therefore no fluxes which can
appear as central charges for particles in space-time.  This is
consistent with the lack of one form gauge fields in space-time.

\newsec{Extensions of this work}

One natural extension of this work is to compactify these theories on
other five dimensional manifolds.  For example, we can compactify our
type II theories on five manifolds, which break half the supersymmetries
like $K3 \times S^1$.  This will extend the work of
\ref\ktso{S. Govindarajan, hep-th/9705113;  M. Berkooz and M. Rozali,
hep-th/9705175.} 
and will give a description of M theory on 
$K3 \times S^1$ (note that these are different $K3$'s).  Here the moduli
space of space-time vacua is $SO(20,4,\IZ) \backslash SO(20,4) /(SO(20)
\times SO(4))$.  It appears as the moduli space of parameters labeling
the compactification of our new theories.  As in the compactification on
$T^5$, this is not just the geometric moduli space but includes the $B$
field and the stringy identifications (mirror symmetry).

We can also try to compactify the heterotic theories on five manifolds,
which break half of the supersymmetries.  This should yield a Matrix
description of compactifications to six dimensions with (1,0)
supersymmetry.

In all the examples we studied the NS5-branes were localized in
$\BR^4$.  This space can be replaced with any consistent string
background.  For example, we can compactify some of it.  If we want to
have an arbitrary number of NS5-branes, we should keep at least 3
non-compact directions.  This leads us to study any of the 5-branes
localized at points in $\BR^3 \times S^1$, thus finding new non-trivial
theories.  It is possible that these theories give the Matrix model
description of compactifications to lower dimensions.

\bigskip\bigskip
\centerline{\bf Acknowledgments}\nobreak

We would like to thank M. Berkooz, S. Shenker and E. Witten 
for useful discussions.   We are particularly indebted to T. Banks for
collaboration during the early stages of this work and for many helpful
comments. This work was supported in part by
DOE grant \#DE-FG02-96ER40559.

\listrefs
\end